%% file: cosine-set2-backgroundmodeling.tex
\RequirePackage{fix-cm}
\documentclass[twocolumn,epjc3]{svjour3}  
\smartqed  
\RequirePackage{graphicx}
\usepackage{multirow}
\usepackage{xcolor}
\usepackage{amsmath}
\usepackage{amssymb}
\usepackage{dblfloatfix} 
\usepackage{hyperref}
\usepackage{microtype} 
\journalname{Eur. Phys. J. C}
\begin{document}
%
\title{Background modeling for dark matter search with 1.7 years of COSINE-100 data}


\author{G.~Adhikari\thanksref{addr14} 
	\and
	E.~Barbosa de Souza\thanksref{addr2}
	\and
	N.~Carlin\thanksref{addr3}
	\and
	J.J.~Choi\thanksref{addr4}
	\and
	S.~Choi\thanksref{addr4}
	\and
	M.~Djamal\thanksref{addr6}
	\and
	A.C.~Ezeribe\thanksref{addr7}
	\and
	L.E.~Fran\c{c}a\thanksref{addr3}
	\and
	C.~Ha\thanksref{addr17}
	\and
	I.S.~Hahn\thanksref{addr9,addr16,addr18}
	\and
	E.J.~Jeon\thanksref{corrauthor1,addr8,addr16}
	\and
	J.H.~Jo\thanksref{addr2}
	\and
	H.W.~Joo\thanksref{addr4}
	\and
	W.G.~Kang\thanksref{addr8}
	\and
	M.~Kauer\thanksref{addr11}
	\and
	H.~Kim\thanksref{addr8}
	\and
	H.J.~Kim\thanksref{addr12}
	\and
	K.W.~Kim\thanksref{addr8}
	\and
	S.H.~Kim\thanksref{addr8}
	\and
	S.K.~Kim\thanksref{addr4}
	\and
	W.K.~Kim\thanksref{addr8,addr16}
	\and
	Y.D.~Kim\thanksref{addr8,addr1,addr16}
	\and
	Y.H.~Kim\thanksref{addr8,addr13,addr16}
	\and
	Y.J.~Ko\thanksref{corrauthor2,addr8}
	\and
	E.K.~Lee\thanksref{addr8}
	\and
	H.~Lee\thanksref{addr8,addr16}
	\and
	H.S.~Lee\thanksref{addr8,addr16}
	\and
	H.Y.~Lee\thanksref{addr8}
	\and
	I.S.~Lee\thanksref{addr8}
	\and
	J.~Lee\thanksref{addr8}
	\and
	J.Y.~Lee\thanksref{addr4}
	\and
	M.H.~Lee\thanksref{addr8,addr16}
	\and
	S.H.~Lee\thanksref{addr16}
	\and
	S.M.~Lee\thanksref{addr8,addr16}
	\and
	D.S.~Leonard\thanksref{addr8}
	\and
	W.A.~Lynch\thanksref{addr7}
	\and
	B.B.~Manzato\thanksref{addr3}
	\and
	R.H.~Maruyama\thanksref{addr2}
	\and
	R.J.~Neal\thanksref{addr7}
	\and
	S.L.~Olsen\thanksref{addr8}
	\and
	B.J.~Park\thanksref{addr8,addr16}
	\and
	H.K.~Park\thanksref{addr15}
	\and
	H.S.~Park\thanksref{addr13}
	\and
	K.S.~Park\thanksref{addr8}
	\and
	R.L.C.~Pitta\thanksref{addr3}
	\and
	H.~Prihtiadi\thanksref{addr6}
	\and
	S.J.~Ra\thanksref{addr8}
	\and
	C.~Rott\thanksref{addr10}
	\and
	K.A.~Shin\thanksref{addr8}
	\and
	A.~Scarff\thanksref{addr7}
	\and
	N.J.C.~Spooner\thanksref{addr7}
	\and
	W.G.~Thompson\thanksref{addr2}
	\and
	L.~Yang\thanksref{addr14}
	\and
	G.H.~Yu\thanksref{addr10} \\	
	(COSINE-100 Collaboration)
}

\thankstext{corrauthor1}{e-mail: ejjeon@ibs.re.kr}
\thankstext{corrauthor2}{e-mail: yjko@ibs.re.kr}


\institute{Center for Underground Physics, Institute for Basic Science (IBS), Daejeon 34126, Republic of Korea \label{addr8}
	\and
	Department of Physics, University of California, San Diego, La Jolla, CA 92093, USA \label{addr14}
	\and
	Department of Physics and Wright Laboratory, Department of Physics, Yale University, New Haven, CT 06520, USA \label{addr2} 
	\and
	Physics Institute, University of S\~{a}o Paulo, 05508-090, S\~{a}o Paulo, Brazil \label{addr3}
	\and
	Department of Physics and Astronomy, Seoul National University, Seoul 08826, Republic of Korea \label{addr4}
	\and
	Department of Physics, Bandung Institute of Technology, Bandung 40132, Indonesia \label{addr6}
	\and
	Department of Physics and Astronomy, University of Sheffield, Sheffield S3 7RH, United Kingdom \label{addr7}
	\and
	Department of Physics, Chung-Ang University, Seoul 06973, Republic of Korea \label{addr17}
	\and
	Department of Science Education, Ewha Womans University, Seoul 03760, Republic of Korea \label{addr9}
	\and
	Center for Exotic Nuclear Studies, Institute for Basic Science (IBS), Daejeon 34126, Republic of Korea \label{addr18}
	\and
	IBS School, University of Science and Technology (UST), Daejeon 34113, Republic of Korea \label{addr16}
	\and
	Department of Physics and Wisconsin IceCube Particle Astrophysics Center,
	University of Wisconsin-Madison, Madison, WI 53706, USA \label{addr11}
	\and
	Department of Physics, Kyungpook National University, Daegu 41566, Republic of Korea \label{addr12}
	\and
	Department of Physics and Astronomy, Sejong University, Seoul 05006, Republic of Korea \label{addr1}
	\and
	Korea Research Institute of Standards and Science, Daejeon 34113, Republic of Korea \label{addr13}
	\and
	Department of Accelerator Science, Graduate School, Korea University, Sejong 30019, Republic of Korea \label{addr15}
	\and
	Department of Physics, Sungkyunkwan University, Seoul 16419, Republic of Korea \label{addr10}
}

\date{Received: date / Accepted: date}

\maketitle

\begin{abstract}
We present a background model for dark matter searches using an array of NaI(Tl) crystals in the COSINE-100 experiment that is located in the Yangyang underground laboratory. The model includes background contributions from both internal and external sources, including cosmogenic radionuclides and surface $^{210}$Pb contamination.
To build the model in the low energy region, with a threshold of 1~keV, we used a depth profile of  $^{210}$Pb contamination in the surface of the NaI(Tl) crystals determined in a comparison between measured and simulated spectra. We also considered the effect of the energy scale errors propagated from the statistical uncertainties and the nonlinear detector response at low energies. 
The 1.7 years COSINE-100 data taken between October 21, 2016 and July 18, 2018 were used for this analysis. 
Our Monte Carlo simulation provides a non-Gaussian peak around 50~keV originating from beta decays of bulk $^{210}$Pb in a good agreement with the measured background. 
This model estimates that the activities of bulk $^{210}$Pb and $^{3}$H are dominating the background rate that amounts to an average level of 2.85$\pm$0.15 counts/day/keV/kg in the energy region of (1--6)~keV, using COSINE-100 data with a total exposure of 97.7~kg$\cdot$years.
\end{abstract}  

\section{Introduction}
\label{intro}
COSINE-100 is an NaI-based experiment for the direct detection of dark matter particles, with an array of 106 kg NaI(Tl) crystals. It has been operating at the Yangyang underground laboratory (Y2L) since September 2016~\cite{cosinedet17,cosinenature18,cosineprl2019}.  One of the COSINE-100 goals is to test DAMA/LIBRA's assertion of an observation of annual modulation signal~\cite{bernabei13,bernabei18}.  
The DAMA/LIBRA collaboration claimed that their results with the release of Phase-II data and 1~keV energy threshold einforces the annual modulation signature at 9.5~$\sigma$ C.L. in the energy region of (1--6)~keV~\cite{bernabei18}.  
There are several groups, such as ANAIS~\cite{amare19}, PICOLON~\cite{picoron}, and SABRE~\cite{sabre19}, using the same low-background NaI(Tl) crystals with the goal of reproducing the DAMA/LIBRA results.

To verify the DAMA/LIBRA modulation signal, a complete understanding of the background energy spectrum is required.
We have developed a background model by performing Monte Carlo (MC) simulations using the Geant4 toolkit~\cite{geant4}. 
We used 1.7 years of COSINE-100 data taken from October 21, 2016 to July 18, 2018 with a 106 kg array of low background NaI(Tl) crystals.
We used the measured spectrum obtained with a threshold of 1~keV electron equivalent energy~\cite{cosine_1keV}.
To build a complete background model we investigated the low-energy contribution from the surface $^{210}$Pb contamination in the NaI(Tl) crystals~\cite{cosine_surf}, in addition to background simulations of radioactive contaminants, such as natural radioisotopes and cosmogenically activated isotopes inside NaI(Tl) and background sources from the exterior of the crystals. 

\input{experiment}

\input{background-sim}
\input{energyscale}
\input{results}

\section{Conclusion} 
\label{conc}
COSINE-100 has been taking data at Y2L from October 21, 2016. We present the background model for the WIMP search during the first 1.7 years of COSINE-100 data with a total exposure of 97.7~kg$\cdot$years. Our previous analysis with 59.5-day data showed that $^{210}$Pb and $^{3}$H produce the dominant contributions in the energy region of (2--6)~keV. As we lowered the threshold to 1 keV, the background modeling was carried out accordingly.
The model includes background contributions from both internal and external sources, including cosmogenic radionuclides and surface $^{210}$Pb contamination.
To build the background model with the energy threshold as low as 1~keV, we used a depth profile of the surface $^{210}$Pb contamination that is provided by the measurement with a test setup at Y2L. 
We also considered the effect of the energy scale errors propagated from the statistical uncertainty and the nonlinear detector response for the simulated spectrum at low energy. 
This improved background model well matches the measured data not only for single-hit events but also for multiple-hit events.
Extrapolating our background model into the Dark matter ROI, we estimate an average background level of 2.85$\pm$0.15 counts/day/keV/kg in the energy region of (1--6) keV for the five crystals dominated by $^{210}$Pb and $^{3}$H.

\section*{Acknowledgments}
\sloppy 
We thank the Korea Hydro and Nuclear Power~(KHNP) Company for providing underground laboratory space at Yangyang. 
This work is supported by: the Institute for Basic Science~(IBS) under project code IBS-R016-A1, 
NRF-2016R1A2B3008343, and NRF-2018R1D1A1B07048941, Republic of Korea; UIUC campus research board, the Alfred P. Sloan Foundation Fellowship, 
NSF Grants Nos. PHY-1151795, PHY-1457995, DGE-1122492, WIPAC, the Wisconsin Alumni Research Foundation, United States; 
STFC Grant ST/ N000277/1 and ST/K001337/1, United Kingdom; and Grant No. 2017/02952-0 FAPESP, CAPES Finance Code 001, and CNPq 131152/2020-3, Brazil.

%
%

\end{document}

%% file: experiment.tex
\section{The COSINE-100 experiment}
\label{sec:2}
\begin{figure*}[ht]
\centering
\begin{tabular}{cc}
\includegraphics[width=0.45\textwidth]{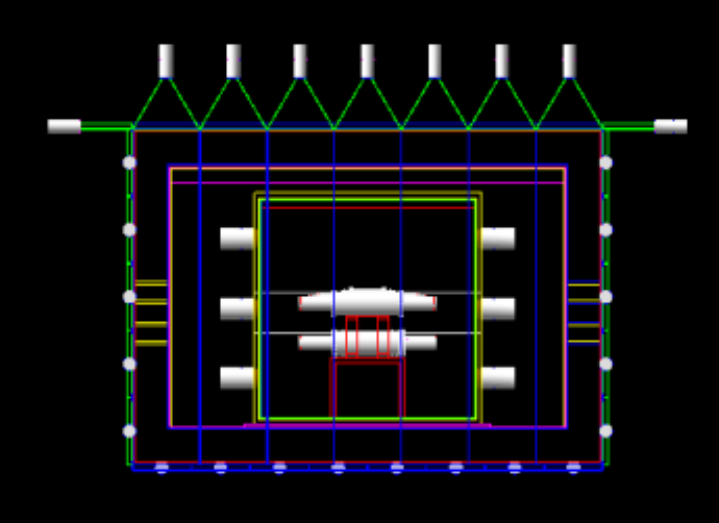} & 
\includegraphics[width=0.35\textwidth]{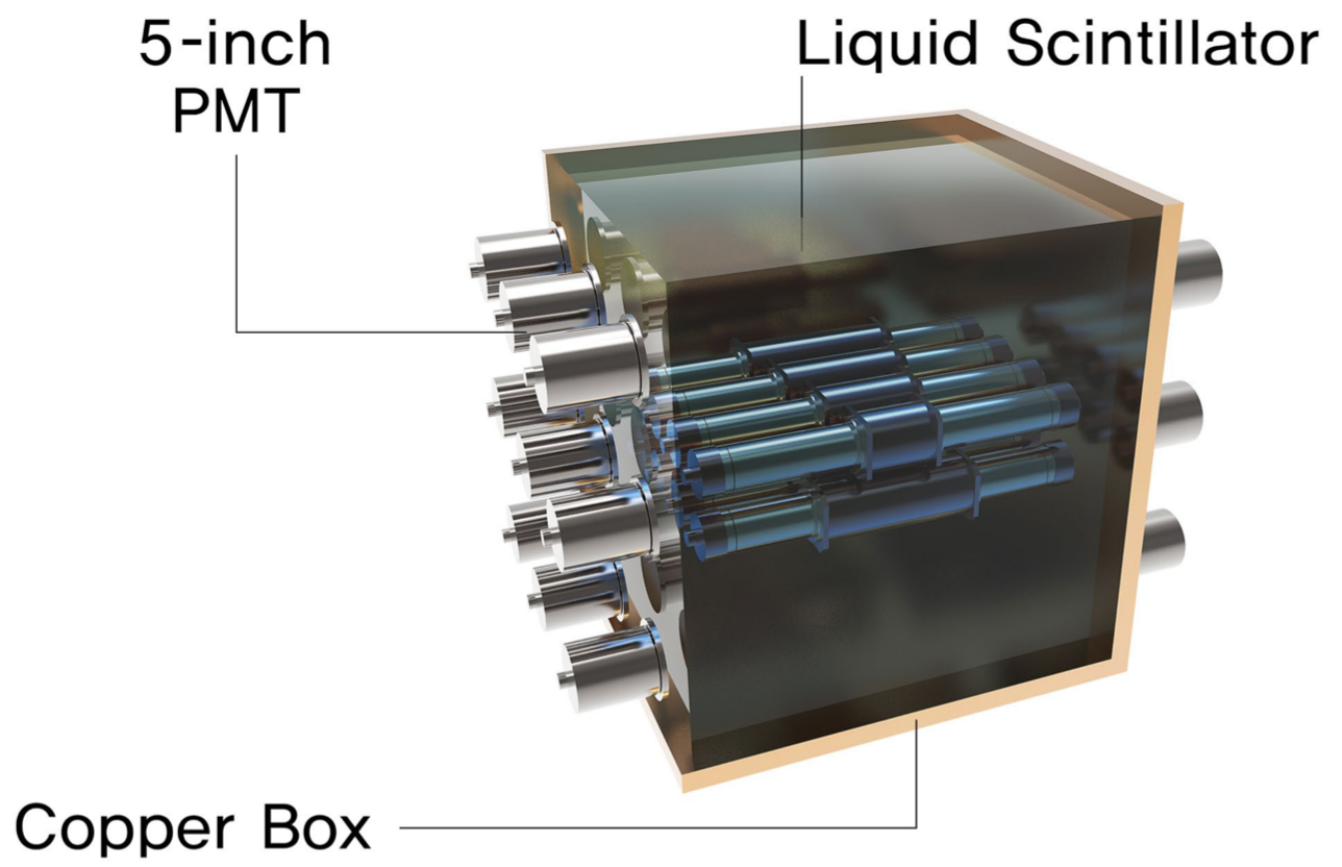} \\
(a) & (b)
\end{tabular}
\caption{Detector geometry (front view) used in the Geant4 simulation. (a) Two white-colored cylindrical shapes inside the center box represent 
the NaI(Tl) detectors supported by the acrylic frame (red) inside the LS. (b) Eighteen 5-inch PMTs are attached to two sides of the copper box to detect LS-produced photons.}
\label{fig:geometry}
\end{figure*}
\begin{figure}[t]
\centering
\includegraphics[width=0.47\textwidth]{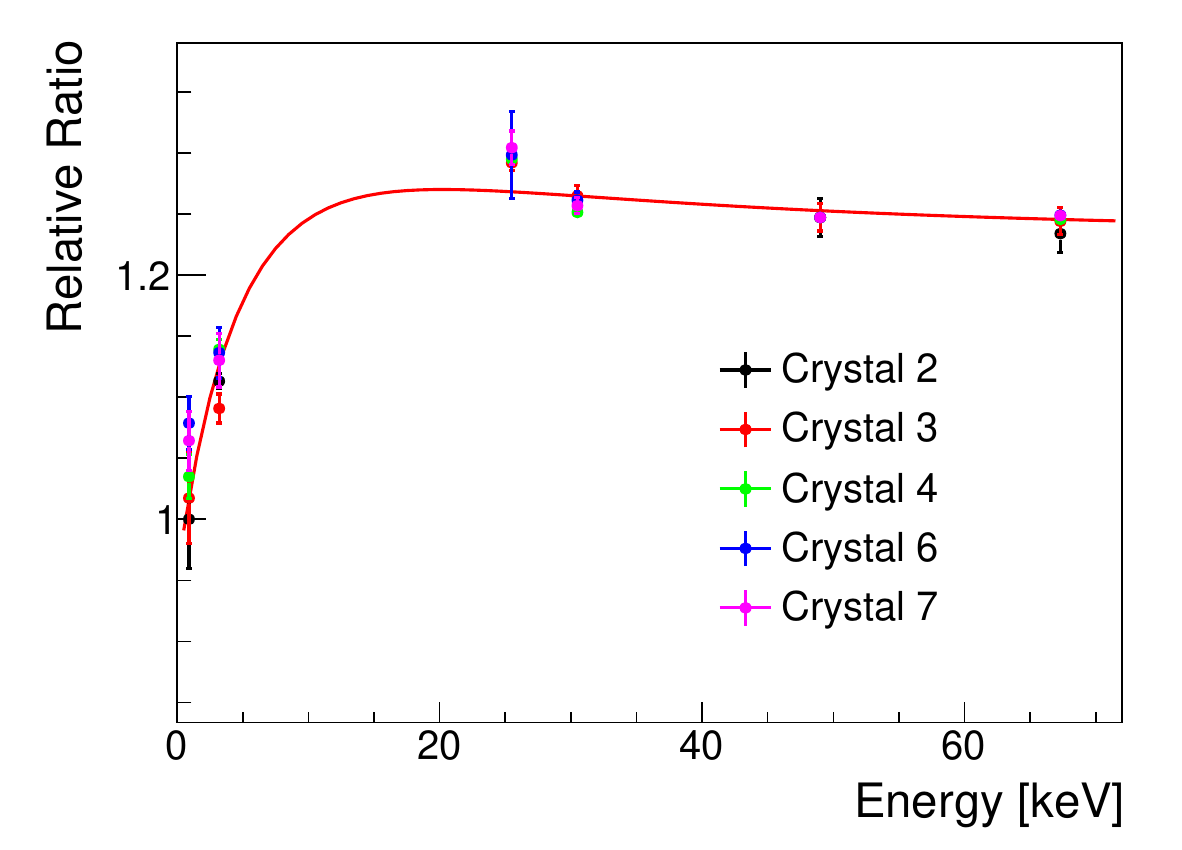}
\caption{Detector response to the calibration data points is described by 
Eq.~\ref{eq:nonlinear} (red line) for five NaI(Tl) crystals at low energy. Data points are $^{22}$Na, $^{40}$K, $^{109}$Cd, $^{121}$Te, $^{210}$Pb, and $^{125}$I from left to right.}
\label{fig:non_cal}
\end{figure}

The main detector of COSINE-100 is a 106 kg array of eight ultra-pure NaI(Tl) crystals (named as C1-C8) stacked in two layers.
Each crystal’s lateral surfaces are wrapped in roughly 10 layers of 250~$\mu$m-thick PTFE reflective sheets~(Teflon) and they are hermetically encased in copper tubes with wall thickness of 1.5~mm and quartz windows (12.0~mm thick) at each end.
The encapsulated NaI(Tl) crystal assembly is equipped with two 3-inch Hamamatsu R12669SEL photomultiplier tubes (PMTs) that are each configured to generate two readouts, a high-gain signal from the anode and a low-gain signal from the 5$^{th}$-stage dynode that are acquired in independent channels. This is because R12669SEL PMTs suffer from non-linear behavior when the signal energy is higher than about 1~MeV and, thus, the dynode signal is amplified with a linear response for energies up to 3~MeV~\cite{cosine_daq}.
The crystals array is immersed in a 2200-liter liquid scintillator (LS) that serves both as an active veto and as a passive shield.
Four shielding layers exist comprising plastic scintillator panels, a lead-brick castle, a copper box, and the tank of LS. Figure~\ref{fig:geometry} shows the detector geometry and the details of the experimental setup is described in Ref.~\cite{cosinedet17}. 

We present here the background modeling to represent 1.7 years of COSINE-100 data with a total effective mass of 61.3~kg from crystals C2, C3, C4, C6 and C7, excluding three crystals due to a high noise rate and low light yields. 
The five crystals used in this analysis have light yields of about 15 photoelectrons/keV and allowed an energy threshold of 2~keV in the previous analysis~\cite{cosinedet17}. 
We consider crystal signals when corresponding to at least four photoelectrons and LS signals when exceeding a 20 keV energy threshold~\cite{cosine_veto}. Crystal events are classified as multiple-hit if they are accompanied by a concomitant signal in one or more other crystals or in the LS. Events for which none of the two occurred are classified as single-hit.

\subsection{Energy calibration}
\label{sec:2.1} 
To determine the light characteristics of the crystals, 
including light yields, energy scales, and energy resolutions, a calibration has been performed using internal $\beta$-- and $\gamma$-ray peaks from radioactive contaminants in the crystals, as well as using $\gamma$-ray sources.
Regarding two readouts from the PMT, the high-gain anode signals are used for low energy events 
up to 70~keV, while the 5$^{th}$-stage dynode signals are used above this energy. 
The peaks at 
238~keV from $^{212}$Pb, 295~keV and 352~keV from $^{214}$Pb, 1173~keV from $^{60}$Co, 1462~keV from $^{40}$K, 2614~keV from $^{208}$Tl, and 609, 1764, and 2204~keV from $^{214}$Bi are used for the high-energy calibration. 
The peaks at 0.9~keV from $^{22}$Na, 3.2~keV from $^{40}$K, 25.5~keV from $^{109}$Cd, 30.5~keV from $^{121}$Te, 49~keV from $^{210}$Pb, and 67.8~keV from $^{125}$I are used for the low-energy calibration; peaks from short-lived cosmogenic isotopes (e.g. $^{125}$I and $^{121}$Te) are obtained using the 59.5 day data. 

In addition, a nonlinear detector response of NaI(Tl) crystals in the low energy region, as reported in Ref.~\cite{non_ref_paper}, is studied and modeled 
empirically as a function of energy, E, as following:  
\begin{equation}
f(E) = p_0\cdot\frac{\ln[(E-p_1)/p_2]}{[(E-p_1)/p_2]^3} + p_3,
\label{eq:nonlinear}
\end{equation}
where $p_{i = 0,1,2,3}$ are free-floating parameters.
The charge-to-energy ratios of the calibration data points for five NaI(Tl) crystals are fitted by Eq.~\ref{eq:nonlinear}, as shown in Fig.~\ref{fig:non_cal}. 

%% file: background-sim.tex
\section{Background simulations}
\label{sec:3}
\begin{figure*}[!b]
\begin{center}
\includegraphics[width=0.95\textwidth]{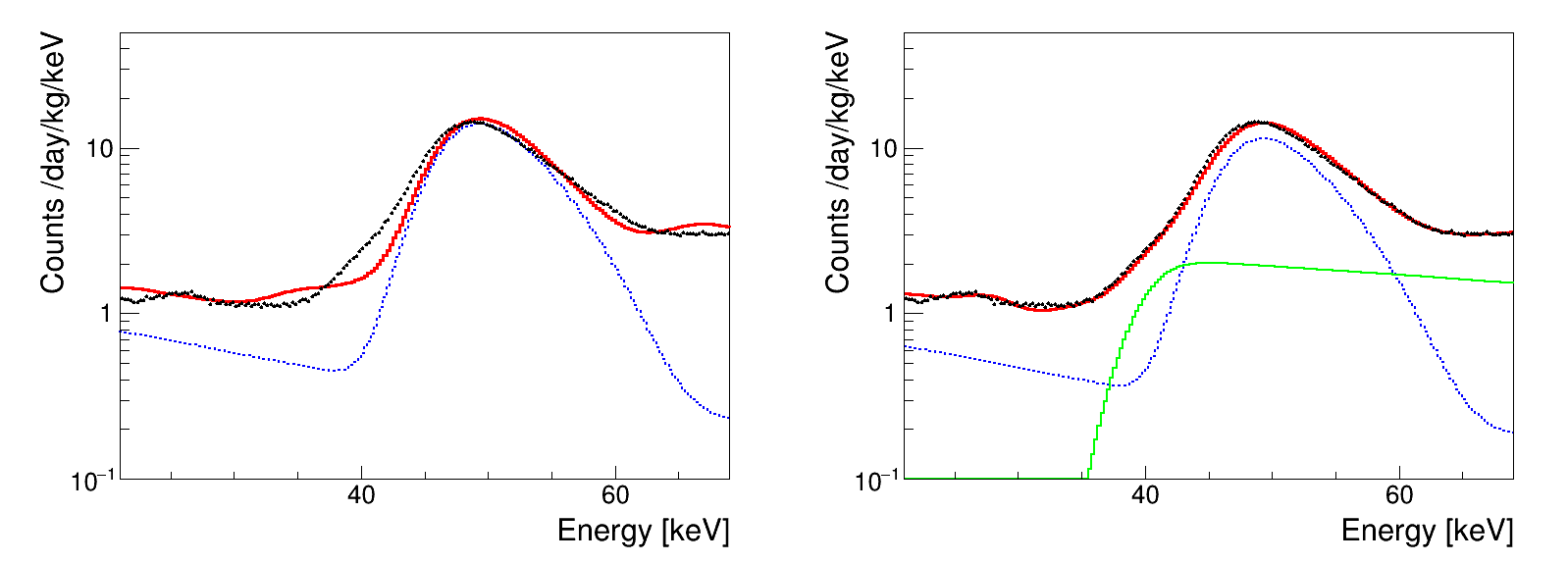} 
\caption{Background modeling for low energy single hit events for Crystal 6. Background models not including $^{129}$I (left) and including $^{129}$I cosmogonic component (right) are shown. The solid green line corresponds to the $^{129}$I isotope. The dashed blue line, the black dots, and the thick red line represent internal $^{210}$Pb isotope, data, and the total MC, respectively.}
\label{model_I129}
\end{center}
\end{figure*}
We use the Geant4-based simulation framework developed for modeling the background spectra of the first 59.5-day of COSINE-100~\cite{cosine_bkg_set1}, although we now adopted G4 version 10.4.2.
The newer version better describes the non-Gaussian peak around 50~keV from the beta decay of $^{210}$Pb; 
it is attributed to 46.5~keV emissions of conversion/Auger electrons and $\gamma$/X-ray together with about 4 keV mean energy of beta electrons from the decay to the excited state of $^{210}$Bi, which results in a non-Gaussian peak. It was not well reproduced by the simulations using G4 version 9.6.2.  

Each simulated event records all energies deposited in the crystals within an event window of 10~$\mu$s from the time when a decay is generated, to account for the conditions in the data acquisition system (DAQ) of the experimental setup [7]. 
Consecutive decays occurring in a short time, such as $^{212}$Bi--$^{212}$Po decays with $^{212}$Po's half-life of 300~ns, appear together in a 10~$\mu$s time window, resulting in pileup events. They are treated as a single event in the simulation. Based on this framework, we carried out MC simulations for all the possible background sources to build a complete model of the background measurements with 1 keV energy threshold.

The simulated spectrum was convolved with the energy resolution as a function of energy obtained during the calibration, described in Sect.~\ref{sec:2.1}.

\subsection{Internal and external backgrounds} 
\label{sec:3.1} 
With the first 59.5 days of data, the detector background was investigated with simulated background spectra. 
We simulated full decay chains of $^{238}$U and $^{232}$Th inside the eight NaI(Tl) crystals assuming a chain equilibrium~\cite{cosine_bkg_set1}.  However, the background levels may vary over the time if the internal activities are not in a chain equilibrium. We indeed found an evident increase in the $^{228}$Th background level during the 1.7 years of data when compared with the 59.5 day data. 
Consequently, the $^{238}$U and $^{232}$Th decay chains are treated as broken at the long-lived parts of the chain in the simulations, i. e. into five groups and three groups, respectively. The activities of each group are free parameters in the fit to the data spectrum.

We simulated external background sources in the COSINE-100 experiment configuration: PMTs, greases, copper cases, bolts, cables, acrylic supports, liquid scintillator, copper box, and a steel structure that supports the lead block housing. 
The background due to the $^{235}$U chain from the PMTs, which was also reported by~\cite{amare16} is included. 
The contribution of $\gamma$s from the $^{208}$Tl decay in $^{232}$Th decay chains in materials outside the shielding as an environmental background is included to identify the peak at 2.614~MeV. 

\subsection{Cosmogenic radioisotopes}
\label{sec:3.2}
Table~\ref{cosmogenic} lists all the cosmogenic radioactive isotopes produced in
the NaI(Tl) crystals in COSINE-100, as reported in Ref.~\cite{cosine_cosmo}, with their half-lives; 
short-lived isotopes, for which half-lives are less than a year, are $^{125}$I, $^{121}$Te, $^{121m}$Te, $^{123m}$Te, $^{125m}$Te, $^{127m}$Te, and $^{113}$Sn and long-lived isotopes are $^{109}$Cd, $^{22}$Na, $^{3}$H, and $^{129}$I. 
There are three long-lived nuclides namely $^{3}$H, $^{22}$Na, and $^{109}$Cd, which have low energy deposits and are, therefore, potentially troublesome. The beta-decay spectrum of tritium has an endpoint energy of 18~keV. The electron capture decay of $^{22}$Na produces 0.87~keV emissions, and the electron capture decay of $^{109}$Cd contributes peaks at 25.5~keV and around 3.5~keV which are at the binding energies of the Ag K-shell and L-shell electrons.
In addition, 
the electron capture decays of $^{113}$Sn and $^{121m}$Te produce an X-ray peak at the L-shell energy of 3~keV.
These short-lived (T$_{1/2}$~$<$~1 year) cosmogenic isotopes are not expected to contribute significantly to the crystals in the long term. However, some backgrounds from them are still expected as averaged activities during 1.7 years.  
It is thus essential to understand their background contributions to the low energy spectra regions, especially in the (1–6) keV dark matter signal region of interest (ROI).
We, therefore, simulated backgrounds from cosmogenic radioactive isotopes, listed in Table~\ref{cosmogenic}. 
The simulated background spectral shapes are used in the data fitting, while the background rates are left free in the fit. The fitted results are compared with the measurements reported in Ref.~\cite{cosine_cosmo} and the details of these comparisons are discussed in Sect.~\ref{sec:results}.
\begin{table}[t]
\begin{center}
\caption{
Cosmogenic radionuclides in the NaI(Tl) crystals identified in other studies and considered here~\cite{cosine_cosmo}. We list the contributing cosmogenic isotopes with their half lives. 
}
\label{cosmogenic}
\begin{tabular}{c|c}
\hline 
Cosmogenic & Half-life~\cite{DDEP}  \\ 
isotopes & (days)  \\ \hline
$^{125}$I & 59.4  \\
$^{121}$Te & 19.17~\cite{Ohya:2010}  \\
$^{121m}$Te & 164.2~\cite{Ohya:2010} \\
$^{123m}$Te & 119.3 \\
$^{125m}$Te & 57.4  \\
$^{127m}$Te & 106.1 \\ 
$^{113}$Sn & 115.1  \\ \hline
$^{109}$Cd & 462 \\
$^{22}$Na & 950  \\
$^{3}$H & 4494  \\
$^{129}$I & 1.57$\times$10$^{7}$ yr~\cite{LNDS} \\ \hline
\end{tabular}
\end{center}
\end{table}

As the presence of cosmogenic $^{129}$I was introduced by DAMA/LIBRA with the estimated concentration of $^{129}$I/$^{nat}$I = (1.7$\pm$0.1)$\times$10$^{-13}$
~\cite{129I-bernabei2008}, we included it in our background fitting model, by treating it as a free parameter. 
The beta decay of $^{129}$I to $^{129}$Xe$^{\ast}$ is followed by $^{129}$Xe$^{\ast}$ transitioning to the stable $^{129}$Xe isotope via the emission of a 39.6~keV $\gamma$--ray and the resulted spectral feature has a distribution 
starting around $\sim$45~keV; this energy region is dominated by bulk $^{210}$Pb that was found to be a main contributor in ROI in the previous analysis with the 59.5 days data. It is thus necessary to distinguish these background contributions, quantitatively. 
Figure~\ref{model_I129} shows the fitted simulation spectra (a) not including $^{129}$I and (b) including $^{129}$I (green solid line).
The inclusion of the contribution from $^{129}$I in Crystal 6, as shown in Fig~\ref{model_I129}(b), 
improves the adherence of our background model to the data around 30 to 70~keV.
The details of the modeling, including $^{129}$I, for each NaI(Tl) crystal are discussed in Sect.~\ref{sec:results}. 

\subsection{Surface $^{210}$Pb}
\label{sec:3.3}
\begin{figure}[t]
\centering
\includegraphics[width=0.45\textwidth, height=0.25\textheight]{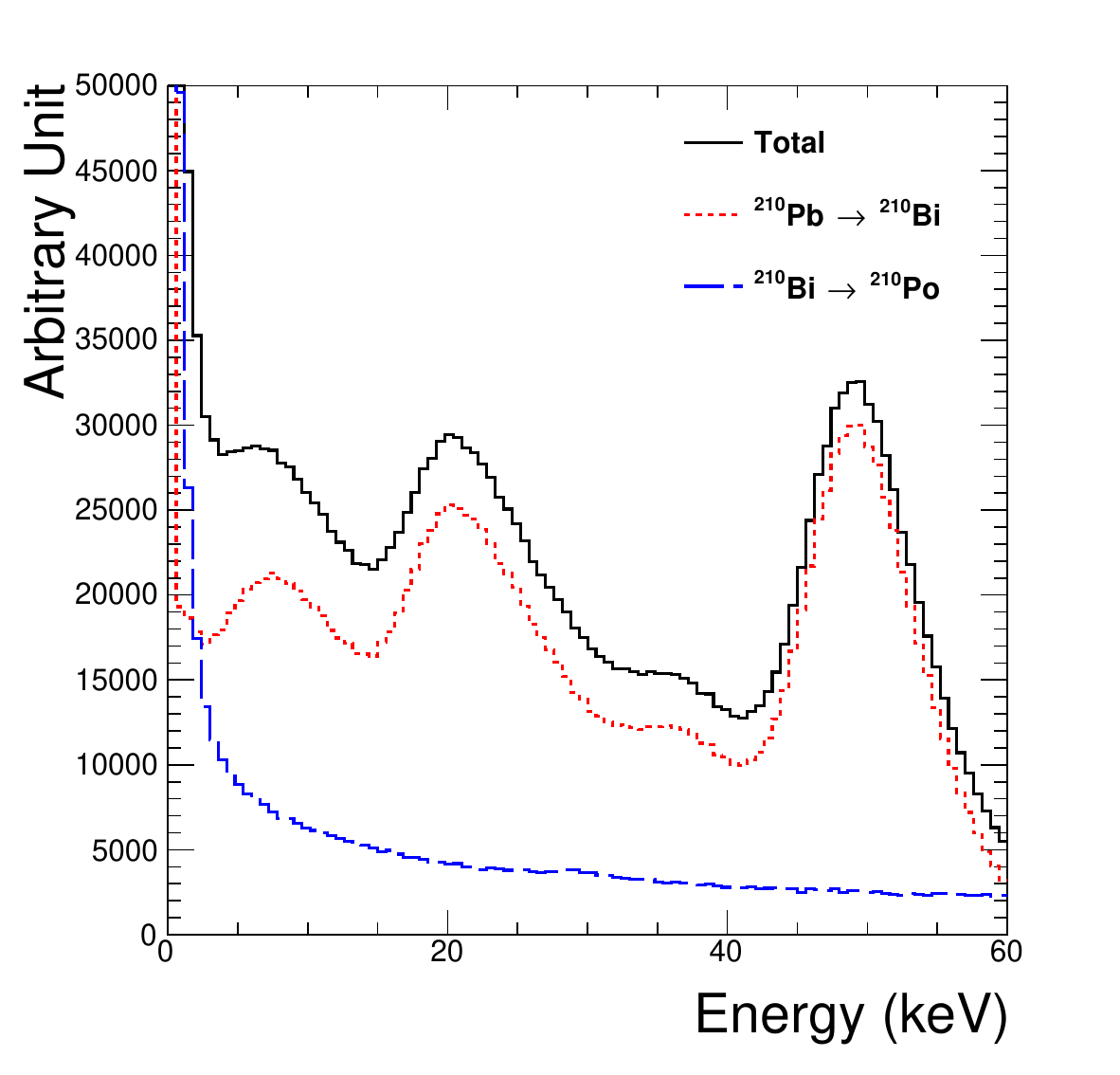}
\caption{
Low-energy spectra due to the beta decays of $^{210}$Pb that are distributed within the surface thickness of 1$\mu$m in the crystal. Further details are given in Sect.~\ref{sec:3.3}.}    
\label{betadecay} 
\end{figure}
The Q value of the beta decay of $^{210}$Pb is 63.5~keV; it decays to the stable $^{210}$Bi with a branching ratio of 16\% and decays to the excited state of $^{210}$Bi at 46.5~keV with a branching ratio of 84\%. Because the de-excitation of the $^{210}$Bi excited state associates with low-energy emissions of electrons and $\gamma$/X--rays, it leaves the full energy deposition with the peak at $\sim$50~keV if it is positioned deep enough in the crystal, while the spectral features of these events depend on the depth distribution of $^{210}$Pb within the crystal surface.   
It has also been suggested that the surface $^{210}$Pb are attributed to the $^{222}$Rn contamination that occurred anytime during the powder- and/or crystal-processing stages.
To understand the energy spectra from the beta decays of $^{210}$Pb in the crystal surface, 
we simulated them by generating $^{210}$Pb at random locations within the surface thickness of 1~$\mu$m in the crystal. The simulated spectra are depicted in Fig.~\ref{betadecay}. 
As expected, there are spectral features in the energy below 40~keV, which are presumably contributed by conversion electrons depending on the depth position of $^{210}$Pb. 
Therefore, the depth profile of the surface $^{210}$Pb contamination should be taken into account for the detector background in the low-energy region of (1--6)~keV. 

%
\begin{figure}[t]
\centering
\includegraphics[width=0.47\textwidth]{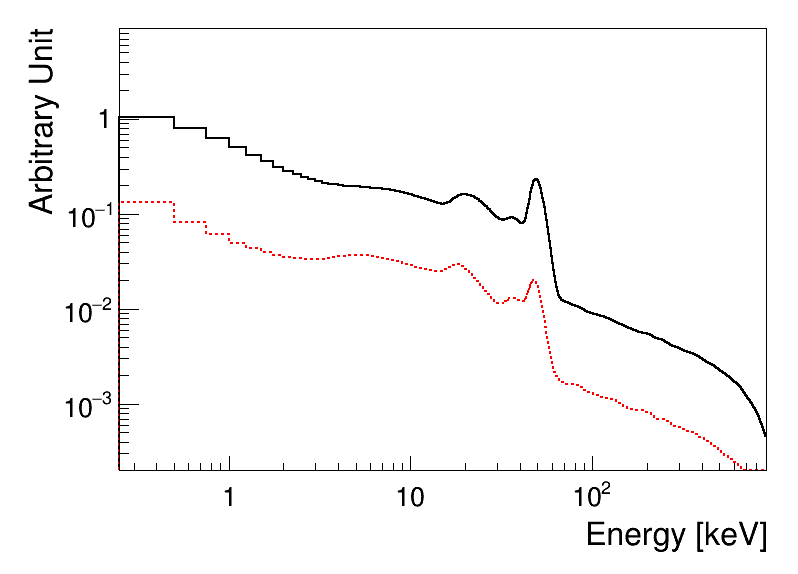}
\caption{Simulated energy spectra for the beta decays of $^{210}$Pb within the surface of Crystal 7; 
we simulated the energy spectra for beta decays of $^{210}$Pb that are exponentially distributed in the surface of Crystal 7 by following two exponential functions with mean depths of 1.39~$\mu$m (black solid line) and 0.107~$\mu$m (red dashed line).
}
\label{surf_prof}
\end{figure}
We have studied the surface $^{210}$Pb contamination with a test setup at Y2L using a NaI(Tl) crystal from the same ingot as C6 and C7~\cite{cosine_surf}. We measured its depth profile by using the measured spectra from both beta decay of $^{210}$Pb and alpha decay of $^{210}$Po of the decay sequence of the surface $^{210}$Pb contamination that is obtained using a $^{222}$Rn-contaminated crystal, as reported in Ref.~\cite{cosine_surf}.
Using this study, it was found that the low-energy spectrum of the surface $^{210}$Pb contamination is primarily attributed to depth profiles of $^{210}$Pb exponentially distributed within a shallow surface with a mean depth of (0.107$\pm$0.003)~$\mu$m, as well as a deep surface with a mean depth of (1.39$\pm$0.02)~$\mu$m. 
We thus simulated beta decays of $^{210}$Pb according to these two depth profiles and included the shapes in the fit to the data. The rates of the two components are left free and independent in the fit as their relative weight might depend on the $^{222}$Rn exposure. Figure~\ref{surf_prof} shows the simulated energy spectra of the two surface components for Crystal 7.

%% file: energyscale.tex
\subsection{Energy scale}
\label{sec:escale}
We have improved the background modeling in the low energy region by precisely studying the low energy contributions from the background sources such as the surface $^{210}$Pb contamination and long-lived cosmogenic isotopes. However, there is still a little mis-matching between data and MC spectra at low energies. It is presumably because the energy scales are set separately for the anode readout and the dynode readout, based on linear fits of calibration data points. They have errors propagated from the statistical uncertainty, as well as the nonlinear detector response, as described in Sect.~\ref{sec:2}. We, thus, consider an adjustment coefficient in the MC spectrum for the energy scale errors.

The energy $E$ in the MC spectrum corresponding to the scaled energy of the dynode readout is adjusted as
\begin{eqnarray}
E\rightarrow E(1+\epsilon),
\label{eq:escale1}
\end{eqnarray}
where $\epsilon$ is a coefficient that represents a change in energy.
The $i^\mathrm{th}$ bin content of the MC spectrum, $B_i$, can be approximated as  
\begin{eqnarray}
B_i \rightarrow B_i + \epsilon\cdot\left.\frac{\partial B_i}{\partial \epsilon}\right|_{\epsilon=0},
\label{eq:escale2}
\end{eqnarray}
where we use a numerical approach to obtain the derivative as
\begin{eqnarray}
\left.\frac{\partial B_i}{\partial \epsilon}\right|_{\epsilon=0}\approx\frac{B(E_i(1+\delta\epsilon)) - B(E_i(1-\delta\epsilon))}{2\delta\epsilon},
\label{eq:escale3}
\end{eqnarray}
where $\delta\epsilon$ represents a very small change in $\epsilon$ and $E_i$ denotes the central value of the $i^{th}$ energy bin. A linear interpolation of the MC spectrum is used for the small variation of $\delta\epsilon$. The coefficient $\epsilon$ in Eq.~\ref{eq:escale2} is determined by fitting the MC spectrum to the data.

\begin{figure}[ht]
\centering
\includegraphics[width=0.45\textwidth]{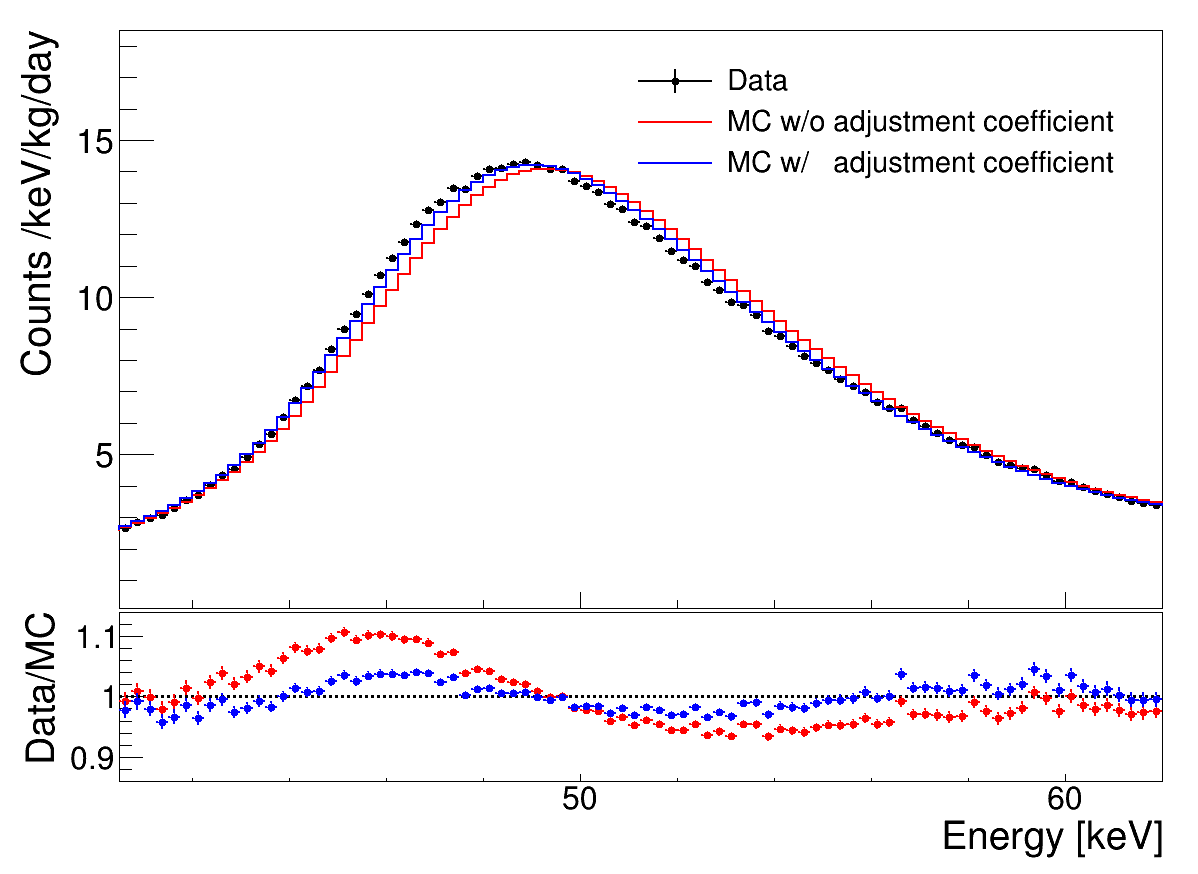}
\caption{
Top panel shows Crystal 6 energy spectra of single- hit events in the low energy region. Black dots are data and the red (blue) line shows the total MC without (with) application of the adjustment coefficient in the background modeling fit. In the bottom panel, the red (blue) dots are the ratio of data to MC without (with) application of the adjustment coefficient.}
\label{fig:anodeES}
\end{figure}

Since there is the nonlinear detector response modeled by the empirical function obtained in Sect.~\ref{sec:2}, at low energies, adjusting the energy in the MC spectrum corresponding to the scaled energy of the anode readout 
follows a different procedure from that of the dynode readout, and is expressed by
\begin{eqnarray}
&E\rightarrow E\left[1+\epsilon\cdot\left\{f(E) - C\right\}\right],
\label{eq:escale4}
\end{eqnarray}
\begin{eqnarray}
B_i \rightarrow B_i + \epsilon\cdot\left\{f(E)-C\right\}\cdot\left.\frac{\partial B_i}{\partial \epsilon}\right|_{\epsilon=0},
\label{eq:escale5}
\end{eqnarray}
where $f(E)$ is the empirical function defined in Eq.~\ref{eq:nonlinear} and $C$ is a coefficient for the linear component determined by fitting the MC spectrum to the data.
Figure~\ref{fig:anodeES} shows the results considering the adjustment coefficient (blue line) and without considering the adjustment coefficient (red line) in the background modeling fit. It is shown that the background modeling has been improved with the adjustment coefficient.

%% file: results.tex
\section{Background modeling and results}
\label{sec:results}
\begin{figure*}[ht]
\begin{center}
\begin{tabular}{cc}
\includegraphics[width=1.0\textwidth]{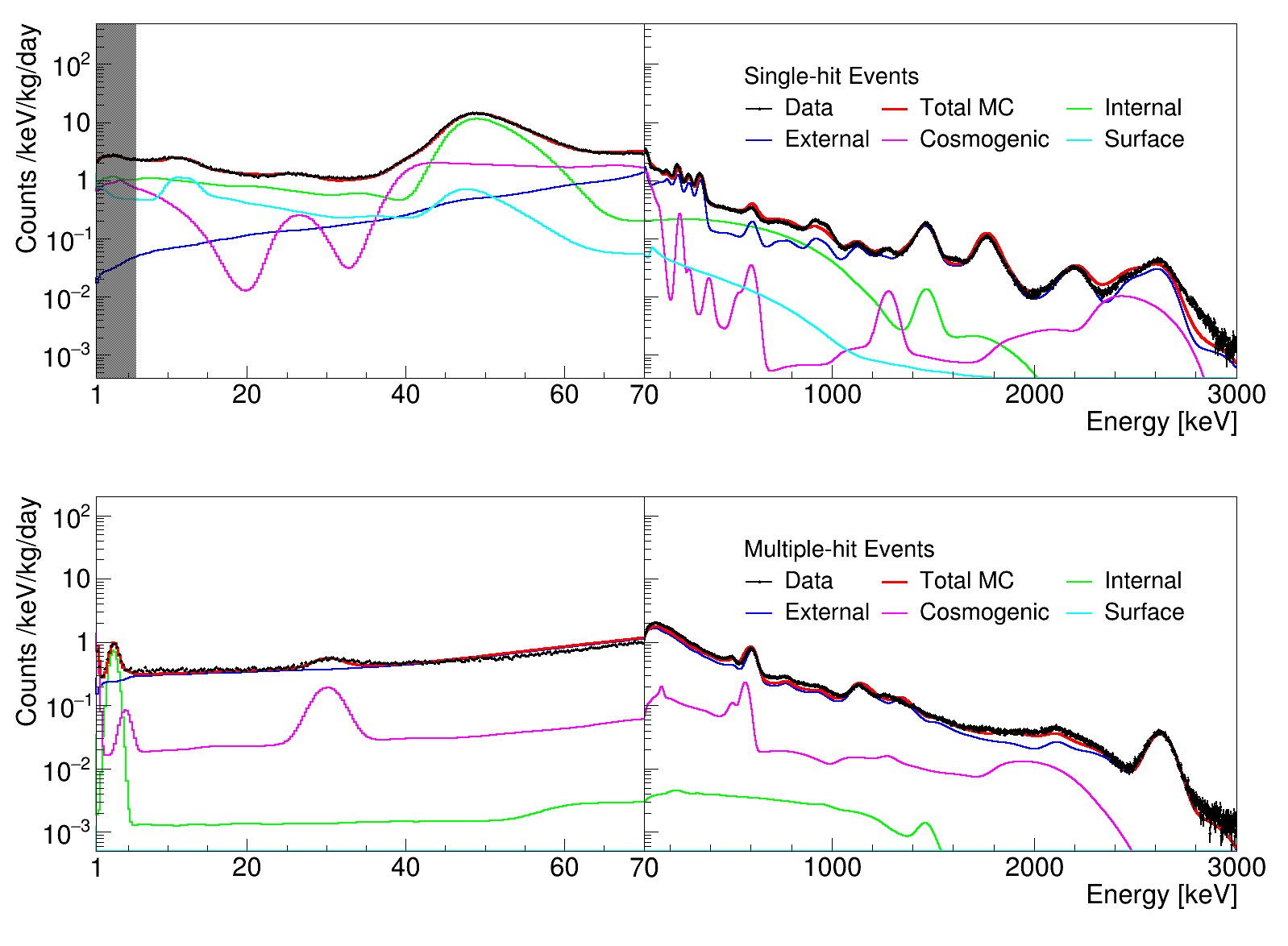}\\
\end{tabular}
\caption{The energy spectra of single-hit (top) and multiple-hit (bottom) events in Crystal~7. 
  The MC was carried out to fit the measured data. The shaded area in the energy spectra of single-hit events is excluded from the data fitting.
  }
\label{fig:bkg_model}
\end{center}
\end{figure*}
\begin{figure*}[ht]
\begin{center}
\begin{tabular}{ccc}
\includegraphics[width=0.45\textwidth]{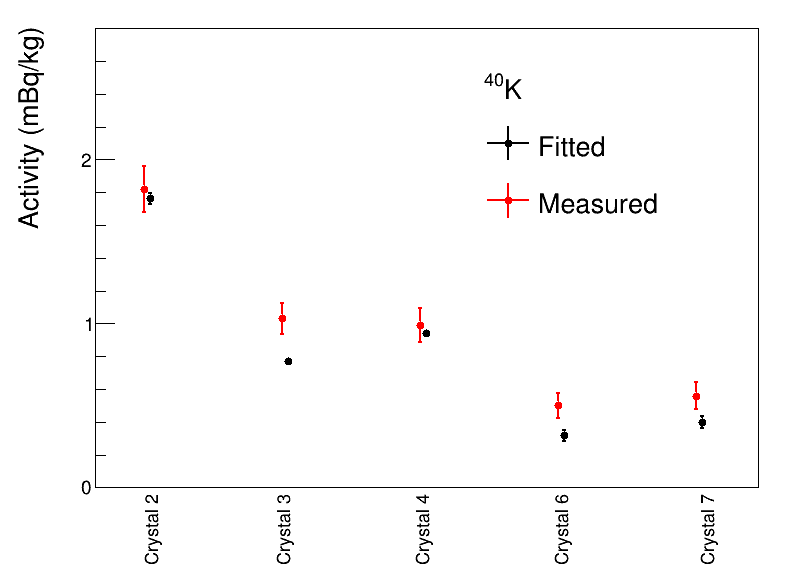} &
\includegraphics[width=0.45\textwidth]{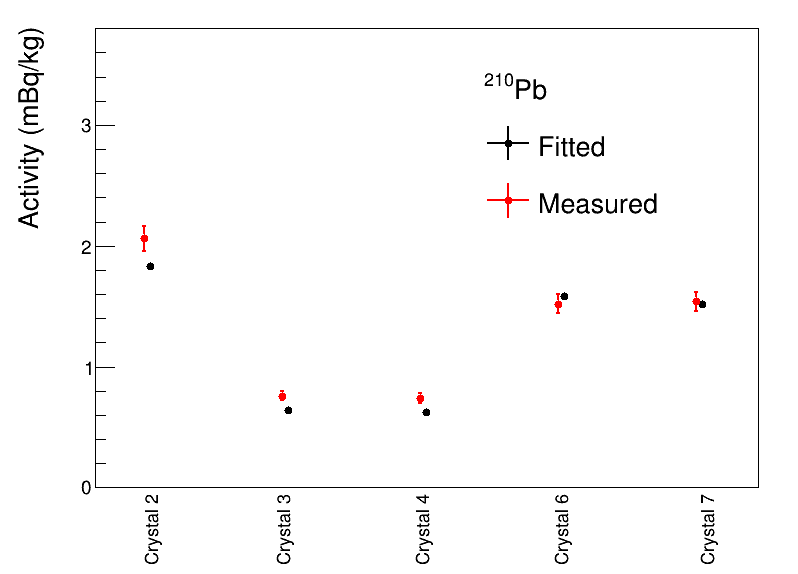} \\
(a) & (b) \\
\includegraphics[width=0.45\textwidth]{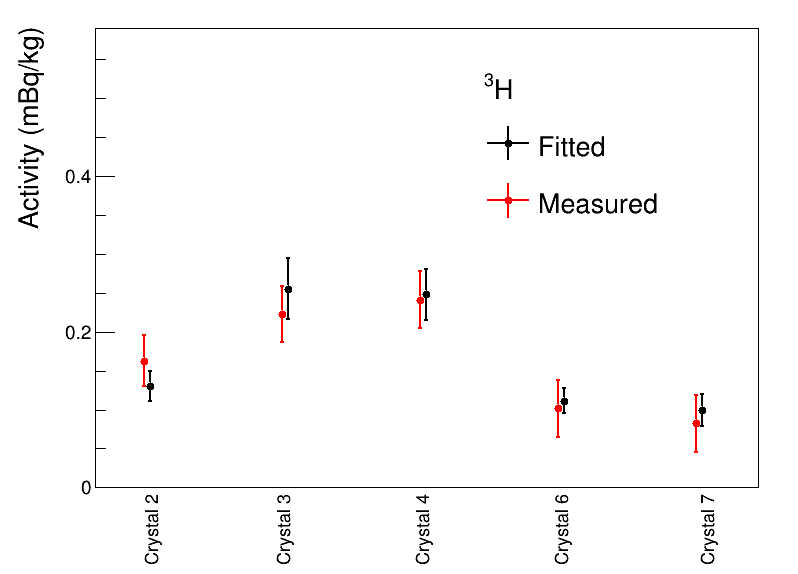} &
\includegraphics[width=0.45\textwidth]{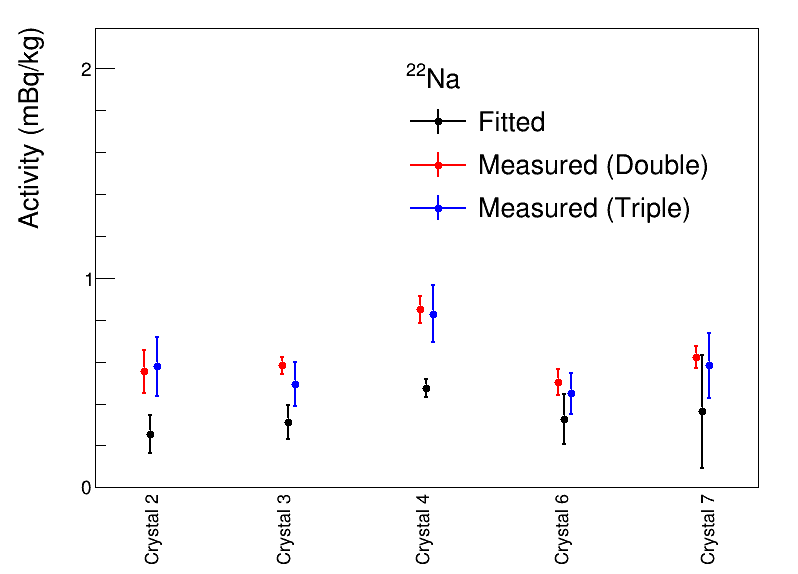} \\
(c) & (d) \\
\includegraphics[width=0.45\textwidth]{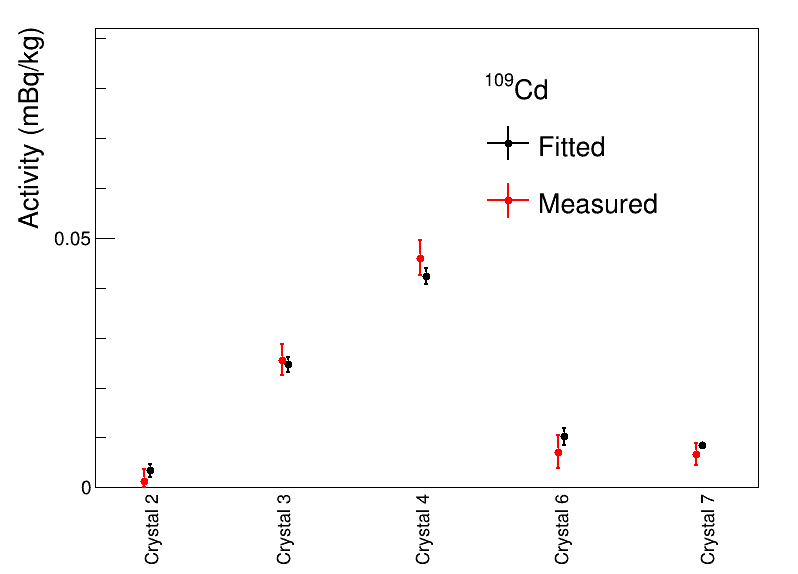} &
\includegraphics[width=0.45\textwidth]{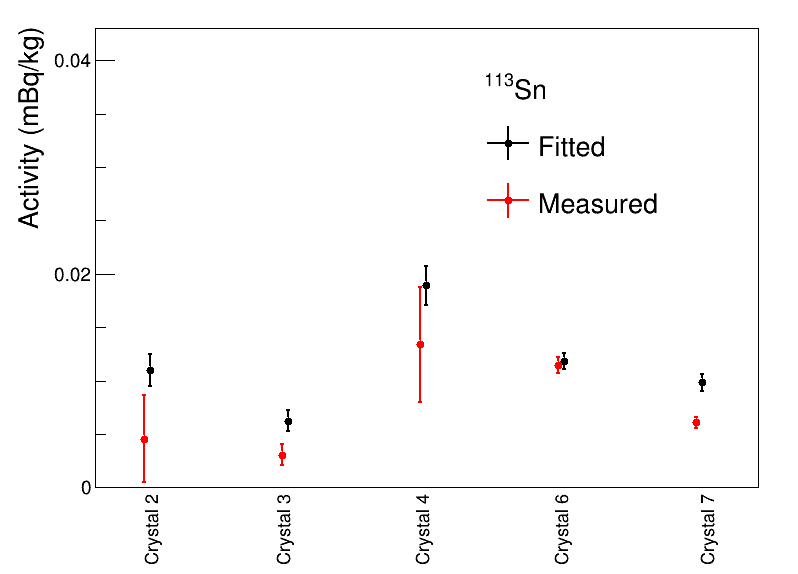} \\
(e) & (f) \\
\end{tabular}
\caption{Comparison of the measured~\cite{cosinedet17}\cite{cosine_bkg_set1}\cite{cosine_cosmo} and the fitted activity levels in five NaI(Tl) crystals. Double and Triple in (d) refer to the double/triple coincidence methods used in Ref.~\cite{cosine_cosmo}.}
\label{fig:comparison}
\end{center}
\end{figure*}
\begin{table*}[ht]
\begin{center}
\caption{Background contributions in the energy range of 1 to 6~keV. There are only statistical uncertainties for the data row, and uncertainties for other rows are from the modeling. In the rows for surface $^{210}$Pb, D and S denote deep and shallow depth, respectively. Corrections for event selection efficiency are applied to the measured data~\cite{cosine_wimp2021}.}
\label{tab:bkg_low}
\begin{tabular}{ccccccc}
\hline
\multicolumn{2}{c}{[Unit: Counts/keV/kg/day]} & Crystal 2 & Crystal 3 & Crystal 4 & Crystal 6 & Crystal 7 \\
\hline
\multicolumn{2}{c}{Data} & $2.834\pm 0.225$ & $3.051\pm 0.482$ & $3.023\pm 0.671$ & $2.458\pm 0.574$ & $2.636\pm 0.421$\\
\hline
\multicolumn{2}{c}{Total simulation} & $2.873\pm 0.193$ & $3.107\pm 0.385$ & $3.077\pm 0.345$ & $2.484\pm 0.347$ & $2.653\pm 0.223$\\
\hline
\multirow{3}{*}{Internal} & $^{210}$Pb & $1.249\pm 0.007$ & $0.434\pm 0.011$ & $0.420\pm 0.012$ & $1.076\pm 0.018$ & $1.053\pm 0.008$\\
                                      & $^{40}$K     & $0.202\pm 0.004$ & $0.083\pm 0.002$ & $0.107\pm 0.001$ & $0.038\pm 0.004$ & $0.047\pm 0.004$\\
                                      & Others         & $0.0104\pm 0.0001 $ & $0.0102\pm 0.0001 $ & $0.0043\pm 0.0002 $ & $0.0074\pm 0.0001 $ & $0.0048\pm 0.0001 $\\
\hline
Surface              & Crystal (D) & $0.182\pm 0.057$ & $0.215\pm 0.147$ & $0.133\pm 0.136$ & $0.195\pm 0.198$ & $0.039\pm 0.088$\\
$^{210}$Pb       & Crystal (S) & $< 0.086$             & $< 0.143$                & $0.114\pm 0.164$ & $0.091\pm 0.250$ & $0.519\pm 0.110$\\
                          & Teflon    & $0.029\pm 0.005$ & $0.083\pm 0.009$ & $0.035\pm 0.008$ & $0.045\pm 0.004$ & $0.054\pm 0.004$\\
\hline
\multirow{4}{*}{Cosmogenic} & $^{3}$H       & $1.091\pm 0.163$ & $2.134\pm 0.326$ & $2.060\pm 0.270$ & $0.929\pm 0.134$ & $0.839\pm 0.172$\\
                                              & $^{113}$Sn & $0.023\pm 0.003$ & $0.013\pm 0.002$ & $0.040\pm 0.004$ & $0.025\pm 0.002$ & $0.021\pm 0.002$\\
                                              & $^{109}$Cd & $0.009\pm 0.003$ & $0.065\pm 0.004$ & $0.113\pm 0.004$ & $0.027\pm 0.004$ & $0.023\pm 0.002$\\
                                              & Others         & $0.023\pm 0.003$ & $0.033\pm 0.001$ & $0.023\pm 0.001$ & $0.016\pm 0.003$ & $0.018\pm 0.006$\\
\hline
\multicolumn{2}{c}{External~($\times10^{-2}$)} & $5.376\pm 0.104$ & $3.669\pm 0.040$ & $2.801\pm 0.039$ & $3.476\pm 0.091$ & $3.509\pm 0.065$\\
\hline
\end{tabular}
\end{center}
\end{table*}
\begin{figure}[ht]
\begin{center}
\begin{tabular}{cc}
\includegraphics[width=0.49\textwidth]{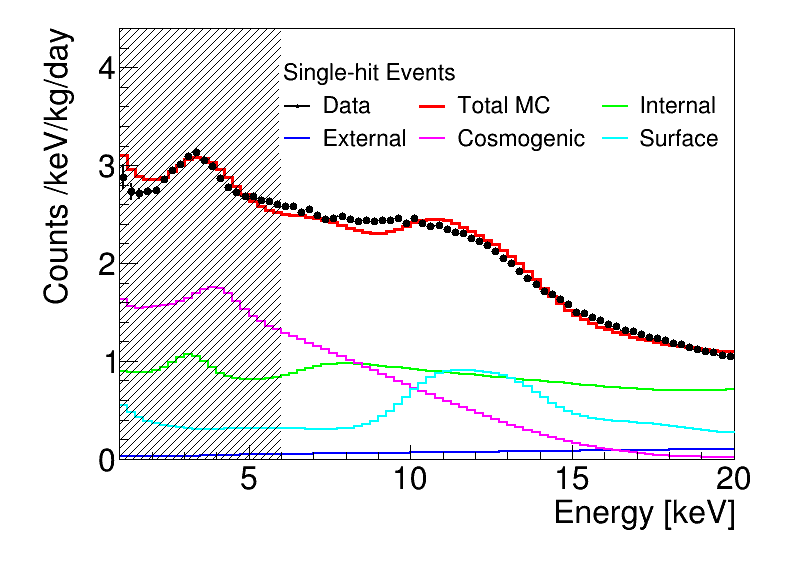}\\ 
\end{tabular}
\caption{
The low-energy spectra of single-hit events averaged for the five crystals.
The measured energy spectrum after efficiency corrections~\cite{cosine_wimp2021} is compared with the total of the simulations. The range of 1 to 6 keV in the MC spectrum is extrapolated from the modeling.
}
\label{fig:bkg_low}
\end{center}
\end{figure}

In order to model the measured energy spectrum ranged from 1~keV quantitatively, we have performed Geant4 MC simulations for the background spectra, as described in Sect.~\ref{sec:3}, which are fitted to the measured data to quantify their background rates. 

We use a binned likelihood method with the following formula~\cite{pdg},
\begin{align}
	-2\ln\lambda(\vec{\alpha}) &= 
		\begin{aligned}[t]
		&2\sum_{i=1}^{N_\mathrm{bins}}\left[\sum_{j=1}^{N_\mathrm{components}}\alpha_jB_{ij} - D_i \right. \\ 
		&\left.+~D_i\ln\frac{D_i}{\sum_{j=1}^{N_\mathrm{components}}\alpha_jB_{ij}}\right] \\		
		&+\sum_{j=1}^{N_\mathrm{components}}\left(\frac{\alpha_j-m_j}{\sigma_j}\right)^2,
		\end{aligned}
\end{align}
where $\lambda(\vec{\alpha})$ is the likelihood ratio in terms of the rates of the MC components $\vec{\alpha}$~=~($\alpha_1$, $\alpha_2$, $\cdots$, $\alpha_{N_\mathrm{components}})$, $D_i$ is the number of events in the $i^\mathrm{th}$ energy bin of the data histogram and $B_{ij}$ is the number of events in the $i^\mathrm{th}$ bin of the $j^\mathrm{th}$ simulation component. The last term denotes a penalty for the rate $\alpha_j$ of the $j^\mathrm{th}$ component and is only active if there is an independent measurement of this component; $m_j$ and $\sigma_j$ are the measured value and the error, respectively. 

As mentioned in Sect.~\ref{sec:escale}, low- and high-energy data are taken through anode and dynode channels, respectively, and they have different energy resolutions. Thus, we perform a four-channel simultaneous fit: single-hit low-energy, single-hit high-energy, multiple-hit low-energy, and multiple-hit high-energy spectra.
The fitting ranges of low-energy spectra are [6,~70] and [1,~70]~keV for single- and multiple-hit events, respectively, while that of high-energy spectra is [70,~3000]~keV for both single- and multiple-hit events. The lower bound of the energy for multiple-hit events is extended to 1~keV based on the study of lowering the energy threshold, reported in Ref.~\cite{cosine_1keV}. The lower bound for the single-hit events is set to 6~keV in order not to bias the WIMP signal in the ROI. 

Figure~\ref{fig:bkg_model}  shows the measured and simulated background spectra of Crystal 7 in both low and high energy regions. 
The spectra for single-hits and multiple-hits are shown in the top and the bottom, respectively.
One can see that the 1.7-year data is well reproduced overall except for the energy region higher than $\sim$2.7~MeV for single-hit events, while it is well reproduced in multiple-hit events. 
This issue is presumed to be due to the absence of one or more components that could better account for the energy range above 2.7 MeV, and the analysis will continue to figure out the issue.
The agreement between the measured and fitted background spectra for both single- and multiple-hit events of Crystal-2, 3, 4, and 6 in both low and high energy regions is as good as shown for Crystal 7.

In the modeling fit, to distinguish between surface and bulk $^{210}$Pb components, 
we used the depth profiles of the surface $^{210}$Pb, obtained in Ref.~\cite{cosine_surf}; there are two depth profiles for shallow and deep depth distributions of $^{210}$Pb in the crystal surface and their fitted results are listed in Table~\ref{tab:bkg_low}.

In Fig.~\ref{fig:comparison}(a) and (b), we compared the fitted activities of internal $^{40}$K and $^{210}$Pb to their measured levels that are determined by the 1.7 years of data for the five crystals with an agreement at the $\sim$20\% level. 
The fitted activities of long-lived cosmogenic isotopes: $^{3}$H, $^{22}$Na, and $^{109}$Cd for the five crystals are compared with the measured ones, reported in Ref~\cite{cosine_cosmo}, as shown in Fig.~\ref{fig:comparison}(c),(d),(e); these values are in reasonable agreement. 
Figure~\ref{fig:comparison}(f) also shows the fitted activities of $^{113}$Sn compared with the measured ones for five crystals.

Based on the background model, we found the background levels for the five NaI(Tl) detectors in unit of counts/day/keV/kg in (1--6)~keV as listed in Table~\ref{tab:bkg_low}; an averaged background level for the five crystals is estimated to be 2.85$\pm$0.15 counts/day/keV/kg in the energy region of (1--6)~keV. 
The dominant background contributions are from $^{210}$Pb and $^{3}$H.
Figure~\ref{fig:bkg_low} shows the low-energy spectra of single-hit events averaged for the five crystals in the (1--20) keV energy region.
The range of 1 to 6 keV in the MC spectrum is extrapolated from the modeling.
The measured energy spectrum after corrections for event selection efficiency~\cite{cosine_wimp2021} is compared with the total of the simulations and is shown to be in a good agreement. 